\documentclass[aps,prl,showpacs]{revtex4}
\usepackage{graphicx}
\usepackage{amsfonts,amssymb,amsmath}
\usepackage{bm}

\begin{document}

\title{A Unified Treatment of Quasi-Exactly Solvable Potentials I}
\date{\today}
 
\author{Ramazan Ko\c{c}}
\email{koc@gantep.edu.tr}
\affiliation{Department of Physics, Faculty of Engineering 
University of Gaziantep,  27310 Gaziantep, Turkey}
\author{Mehmet Koca}
\email{kocam@squ.edu.om}
\affiliation{Department of Physics, College of Science,
Sultan Qaboos University, PO Box 36  \\
Al-Khod 123, Sultanete of Oman}

\begin{abstract}
A set of quasi-exactly solvable quantum mechanical potentials associated
with the P\"{o}schl-Teller potential, the generalized P\"{o}schl-Teller
potential, the Scarf potential, and the harmonic oscillator potential have
been studied. Solutions of the Schr\"{o}dinger equation for each potential
have been determined and the eigenstates are expressed in terms of the
orthogonal polynomials. The potentials are related to each other by suitable
change of variables.
\end{abstract}
\maketitle

\section{Introduction}

The discovery of new class of physically significant spectral problems,
called quasi-exactly solvable(QES) models\cite{Turbiner1,Turbiner2}, has
attracted much attention\cite{Gonzalez, Debergh}. Several methods\cite%
{Kuliy, Brihaye} for the generation of the QES potentials have been worked
out and consequently a number of QES potentials have been suggested. One of
the methods for the calculation of eigenstates and eigenvalues of the QES
potentials is the use of orthogonal polynomials. Bender and Dunne have showed%
\cite{Bender1} that there is a correspondence between the QES models in
quantum mechanics and the set of the orthogonal polynomials $P_{m}(E)$,
which are polynomials in energy $E$. In particular they have demonstrated
the properties of the polynomials for the QES sextic oscillator potential%
\cite{Bender2}. Other aspects of the $P_{m}(E)$ have been discussed in
various papers\cite{Finkel, Krajewska}. In this paper we show that the
polynomial $P_{m}(E)$ normally appears as a factor of the coefficient in the
expansion of the eigenstates of the Schr\"{o}dinger equation in the presence
of the QES P\"{o}schl-Teller potential, the Generalized QES P\"{o}%
schl-Teller potential and the PT symmetric Scarf potential. We also show
that the $P_{m}(E)$ \ satisfies the same three-term recurrence relations for
all three potentials.

An algorithm generating the analytical eigenfunctions as well as the
eigenvalues of the Schr\"{o}dinger equation for various QES potentials is
developed. The procedure presented here reproduces the results of the
exactly solvable Schr\"{o}dinger equations in proper limits. Our formalism
signals that, there is a closer similarity between QES P\"{o}schl-Teller
potential and the perturbed P\"{o}schl-Teller potential\cite{Znojil1,Znojil2}%
. It is well known that the harmonic oscillator and the P\"{o}schl-Teller
potential possess the similar theoretical behavior in a certain range of
variable. In contrast to enormous interest in the perturbed harmonic
oscillator, there have been only a number of studies on the perturbed P\"{o}%
schl-Teller potential\cite{Znojil1}

We show that the QES P\"{o}schl-Teller potential can be transformed on to
the generalized QES P\"{o}schl-Teller potential and the QES PT symmetric
Scarf potential by replacing coordinate $x\rightarrow x/2$ and $x\rightarrow
x/2+i\pi /4,$ respectively. These potentials are also related to the QES
harmonic oscillator potential by redefining the parameters and applying an
appropriate limiting procedure.

The paper is organized as follows. In section 2 we discuss the generation
and solution of the QES P\"{o}schl-Teller potential. The QES generalized P%
\"{o}schl-Teller potential and the QES PT symmetric Scarf potentials are
obtained from the QES P\"{o}schl-Teller potential in sections 3 and 4,
respectively. Transformations of QES P\"{o}schl-Teller potential to the
sextic oscillator potential and the QES PT symmetric Scarf potential to the
PT symmetric anharmonic oscillator potential are discussed in sections 5 and
6 respectively. In conclusive remarks we discuss the implementation of our
method on the other QES potentials which will be the topic of another
publication.

\section{QES P\"{o}schl-Teller Potential}

The QES P\"{o}schl-Teller potential can be generated by several methods\cite%
{Turbiner3, Gonzalez, Znojil1}. One method is to use the Lie algebraic
technique. The linear and bilinear combinations of the operators of the
sl(2,R)\footnote{%
The QES P\"{o}schl-Teller potential studied in this paper can be obtained by
transforming the following linear and bilinear combinations of the operators
of the sl(2,R) Lie algebra, \textit{J}$_{+}$\textit{J}$_{-}$\textit{+J}$_{-}$%
\textit{J}$_{0}$\textit{+(L+j+}$\frac{1}{2}$\textit{)J}$_{-}$\textit{+(B+2j)J%
}$_{0}$\textit{+qA}$^{2} $\textit{J}$_{+}$\textit{-(}$\lambda $\textit{%
-j(B+2j))=0} to the Schr\"{o}dinger equation with standard realization\cite%
{Turbiner3, Gonzalez, Debergh}} Lie algebra with the standard realization%
\cite{Turbiner3, Gonzalez, Debergh}, leads to the following differential
equation, 
\begin{equation}
z(1-z)\frac{d^{2}\Re _{j}(z)}{dz^{2}}+(L+\frac{3}{2}+z(B+4j-qA^{2}z))\frac{%
d\Re _{j}(z)}{dz}-(\lambda -2jqA^{2}z)\Re _{j}(z)=0  \label{eq:1}
\end{equation}%
where $L$, $q,$ $A$ and $\lambda $ are constants and $j=0,1/2,1,\cdots $.
The differential equation(\ref{eq:1}) becomes QES, provided B is taken as: 
\begin{equation}
B=-\frac{1}{2}\left( 2L+8j+5-\sqrt{1+4A(A+1+(2L+8j+5)qA)}\right) .
\label{eq:2}
\end{equation}%
Moreover, it becomes exactly solvable with the condition $q=0$. The function 
$\Re _{j}(z)$ is a polynomial of degree $2j$. In order to obtain the QES
quantum mechanical potentials we can transform (\ref{eq:1}) in the form of
Schr\"{o}dinger equation by introducing the variable 
\begin{equation}
z=-\sinh ^{2}\alpha x.  \label{eq:3}
\end{equation}%
Now we define the wave function as follows: 
\begin{equation}
\psi (x)=(\cosh \alpha x)^{qA^{2}-B-L-4j-2}(\sinh \alpha x)^{1+L}e^{-\frac{1%
}{4}qA^{2}\cosh 2\alpha x}\Re _{j}(-\sinh ^{2}\alpha x).  \label{eq:4}
\end{equation}%
Substituting (\ref{eq:3}) and (\ref{eq:4}) in (\ref{eq:1}) we obtain the Schr%
\"{o}dinger equation with $(\hbar =2m=1)$: 
\begin{equation}
-\frac{d^{2}\psi (x)}{dx^{2}}+(V(x)-E)\psi (x)=0  \label{eq:5}
\end{equation}%
where the potential $V(x)$ is given by

\begin{eqnarray}
V(x) &=&L(L+1)\alpha ^{2}\csc h^{2}\alpha x-A(A+1)\alpha ^{2}\sec
h^{2}\alpha x+  \notag \\
&&q(2BA^{2}\sinh ^{2}\alpha x+qA^{4}\sinh ^{4}\alpha x)\alpha ^{2}\tanh
^{2}\alpha x  \label{eq:6}
\end{eqnarray}%
The eigenvalues of energy are given by the expression%
\begin{equation}
E=\left[ -(L-A+2m+1)^{2}+(L-A+B+4j+2)(2L+4m+3)+4\lambda \right] \alpha ^{2}.
\label{eq:7}
\end{equation}%
One can check that for $q=0$ the potential given in (\ref{eq:6}) is exactly
solvable P\"{o}schl-Teller potential and the eigenstates of the Schr\"{o}%
dinger equation can be expressed in terms of the Jacobi polynomials. It is
easy to see that the equalities $B=A-L-4j-2$ and $\lambda =4m(L-M+m+1)$ \
hold when $q=0$. Then the eigenvalues of the Schr\"{o}dinger equation take
the form 
\begin{equation}
E=-\alpha ^{2}\left[ (L-A+2m+1)^{2}+4m(L-M+m+1)\right] .
\end{equation}

Next task is now to determine the eigenfunction of the QES Schr\"{o}dinger
equation(\ref{eq:5}). Therefore, we search for a solution of (\ref{eq:1}) by
substituting the polynomial%
\begin{equation}
\Re _{j}(z)=\sum\limits_{m=0}^{2j}a_{m}z^{m}  \label{eq:9}
\end{equation}%
which leads to an expression for the coefficients $a_{m}$%
\begin{equation}
a_{m}=\frac{(4qA^{2})^{m}(2j)!(2L+1)!(L+m)!}{2m!(2j-m)!(2L+1+2m)!}%
P_{m}(\lambda ).
\end{equation}%
The polynomial $P_{m}(\lambda )$ satisfies the following three-term
recurrence relation%
\begin{equation}
2(2j-m)qA^{2}P_{m+1}(\lambda )+m(2L+2m+1)P_{m-1}(\lambda )-2(\lambda
+m(B+4j-m+1))P_{m}(\lambda )=0  \label{eq:10}
\end{equation}%
with the initial condition $P_{0}(\lambda )=1.$ The polynomial $%
P_{m}(\lambda )$ vanishes for $m\geqslant 2j+1$ and the roots of $%
P_{2j+1}(\lambda )=0$ corresponds to the $\lambda -$eigenvalues of the Schr%
\"{o}dinger equation(\ref{eq:5}). The first three of these polynomials are
given by 
\begin{eqnarray}
P_{1} &=&{}\lambda  \notag \\
P_{2} &=&{}\lambda ^{2}-(B+4j)\lambda -j(2L+3)qA^{2}  \notag \\
P_{3} &=&{}\lambda ^{3}-(3B12j-4)\lambda ^{2}+  \label{eq:11} \\
&&\left[ 2B(B-2)+16j(B+2j-1)+(2L-j(6L+13)+5)qA^{2}\right] \lambda +  \notag
\\
&&2j(2L+3)(B+4j-2)qA^{2}  \notag
\end{eqnarray}

The recurrence relation (\ref{eq:10})can also be put in the matrix form. The
tridiagonal matrix characterizes the system,%
\begin{equation}
\left( 
\begin{array}{ccccc}
\beta _{0}-4\lambda & \mu _{_{2j}} &  &  &  \\ 
\gamma _{1} & \beta _{1}-4\lambda & \mu _{_{2j}-1} &  &  \\ 
& \ddots & \ddots & \ddots &  \\ 
&  & \gamma _{2j-1} & \beta _{2j-1}-4\lambda & \mu _{_{1}} \\ 
&  &  & \gamma _{2j} & \beta _{2j}-4\lambda%
\end{array}%
\right) \left( 
\begin{array}{c}
P_{0} \\ 
P_{1} \\ 
\vdots \\ 
P_{2j-1} \\ 
P_{2j}%
\end{array}%
\right) =0  \label{eq:12}
\end{equation}%
where the parameters in matrix elements are given by%
\begin{equation}
\gamma _{m}=2m(2L+2m+1),\quad \mu _{m}=4mqA^{2},\quad and\quad \beta
_{m}=4m(B+4j-m+1).
\end{equation}%
Analytical solutions of the recurrence relation (\ref{eq:10}) and the
determinant of (\ref{eq:12}) for $\lambda $ are available only for the first
few values of $j\leqslant 2$. For $j>2$ the solutions become numerical, and
the numerical errors grow rapidly. The solutions take simpler forms and the
precision becomes better for $A\gg q,$ in which case B takes the value 
\begin{equation}
B\approx 2(\lambda +m(L-A+m+1))-m(2L+8j+5)qA.
\end{equation}%
In this approximation the QES P\"{o}schl-Teller potential in (\ref{eq:7})
becomes comparable with the perturbed P\"{o}schl-Teller potential, a
specific form of which has been studied in \cite{Znojil3}.

\section{Generalized QES P\"{o}schl-Teller Potential}

In this section we present a procedure that relates the QES P\"{o}%
schl-Teller potential to the \textit{generalized} QES P\"{o}schl-Teller
potential. It is amusing to observe that when the coordinate $x$ of the P%
\"{o}schl-Teller potential is replaced by $x\rightarrow x/2$, the QES P\"{o}%
schl-Teller potential transforms to the generalized QES P\"{o}schl-Teller
potential:%
\begin{eqnarray}
V(x) &=&\frac{\alpha ^{2}}{2}[\left( L(L+1)+A(A+1)\right) \csc h^{2}\alpha x
\notag \\
&&+(L-A)(L+A+1)\coth \alpha x\csc h\alpha x]  \label{eq:13} \\
&&+q(2BA^{2}\sinh ^{2}\frac{\alpha x}{2}+qA^{4}\sinh ^{4}\frac{\alpha x}{2}%
)\alpha ^{2}\tanh ^{2}\frac{\alpha x}{2}  \notag
\end{eqnarray}%
Obviously when $q=0$ this potential reduces to the exactly solvable
generalized P\"{o}schl-Teller potential. The wave function corresponding to
the solution of (\ref{eq:5}) with the potential of (\ref{eq:13}) now takes
the form%
\begin{equation}
\psi (x)=(\cosh \frac{\alpha x}{2})^{qA^{2}-B-L-4j-2}(\sinh \frac{\alpha x}{2%
})^{1+L}e^{-\frac{1}{4}qA^{2}\cosh \alpha x}\Re _{j}(-\sinh ^{2}\frac{\alpha
x}{2}).  \label{eq:14}
\end{equation}%
The corresponding energies are determined as%
\begin{equation}
E^{\prime }=\frac{E}{4}+m(B+4j-m+1)\alpha ^{2}  \label{eq:15}
\end{equation}%
where $E$ is the eigenvalue of the Schr\"{o}dinger equation with the QES P%
\"{o}schl-Teller potential given by(\ref{eq:7}). It is clear that the same
recurrence relation in(\ref{eq:10}) for the polynomial $P_{m}(\lambda )$
holds true.

\section{QES PT\ Symmetric Scarf Potential}

It is interesting to observe that one can transform the QES P\"{o}%
schle-Teller potential to the PT symmetric QES Scarf potential by replacing $%
x$ by the complex variable%
\begin{equation}
x\rightarrow \frac{x}{2}+\frac{i\pi }{4\alpha }
\end{equation}%
Then one obtains the following potential:\ 
\begin{align}
V(x)& ={}-\frac{\alpha ^{2}}{2}[\left( L(L+1)+A(A+1)\right) \sec h^{2}\alpha
x  \notag \\
& +i(L-A)(L+A+1)\tanh \alpha x\sec h\alpha x]  \label{eq:16} \\
& +\frac{\alpha ^{2}}{4}(2qBA^{2}\sinh ^{2}(\frac{\alpha x}{2}+\frac{i\pi }{4%
})+q^{2}A^{4}\sinh ^{4}(\frac{\alpha x}{2}+\frac{i\pi }{4}))\tanh ^{2}(\frac{%
\alpha x}{2}+\frac{i\pi }{4})  \notag
\end{align}%
This is the QES form of the PT symmetric Scarf potential which has not been
discussed in the literature. The exactly solvable part of the potential (\ref%
{eq:16})(given in the [...] which, can be obtained if $q=0$) has been
discussed recently by Bagchi\cite{Bagchi}. The energy eigenvalues are the
same as in (\ref{eq:15}) obtained for the generalized QES P\"{o}schle-Teller
potential. But the wave function now reads

\begin{eqnarray}
\psi (x) &=&{}\left( \frac{i+\tanh \alpha x}{i-\tanh \alpha x}\right) ^{%
\frac{1}{2}(B+2L-qA^{2}+4j+3)}(\cosh \alpha x)^{\frac{1}{2}%
(qA^{2}-4j-B-1)}\times  \notag \\
&&{}e^{-\frac{i}{4}qA^{2}\sinh \alpha x}\Re _{j}(-\sinh (\frac{\alpha x}{2}+%
\frac{i\pi }{4}).  \label{eq:17}
\end{eqnarray}
Here again the polynomial $P_{m}(\lambda )$ satisfies the same recurrence
relation defined in(\ref{eq:10})

The method described in this section in order to transform one type of
potential to another type may be further generalized. The transformation $%
x\rightarrow ax+ib$ , where $a$ and $b,$ are arbitrary real parameters,
preserves the PT symmetry. The energy eigenvalues of the new potential
involves further terms in addition to the scaled energy eigenvalues of the
QES P\"{o}schl-Teller potential. The corresponding wave function is obtained
from the former wave function of the P\"{o}schl-Teller potential by
replacing $x\rightarrow ax+ib,$ and the polynomial $P_{m}(\lambda )$ obeys
the same recurrence relation of(\ref{eq:10}). This transformation has been
discussed in the literature\cite{Znojil3} for the exactly solvable
potentials.

\section{The Sextic oscillator}

In this section we discuss a method about the transformation of the QES P%
\"{o}schle-Teller potential to the sextic oscillator potential. The QES P%
\"{o}schle-Teller potential can be converted to the radial sextic oscillator
potential by redefining the parameters and taking suitable limits while
keeping the variable $x$ intact. If we redefine the parameters in (\ref{eq:6}%
) and (\ref{eq:7}) by introducing%
\begin{equation}
B=\frac{b}{\alpha ^{2}},\quad A=\frac{a}{\alpha ^{2}},\quad \lambda =\frac{%
\varepsilon -mb}{\alpha ^{2}}
\end{equation}%
and taking the limit of the $(V(x)-E)$ term in the Schr\"{o}dinger equation
for $\alpha \rightarrow 0$ the P\"{o}schle-Teller potential transforms to
the radial sextic harmonic oscillator potential:%
\begin{equation}
V=\frac{L(L+1)}{x^{2}}%
+(b^{2}-(2L+8j+5)qa^{2})x^{2}+2bqa^{2}x^{4}+q^{2}a^{4}x^{6}.  \label{eq:18}
\end{equation}%
The ground state wave function(\ref{eq:4}) takes the form%
\begin{equation}
\psi (x)=x^{1+L}e^{-\frac{b}{2}x^{2}-\frac{q}{4}x^{4}}
\end{equation}%
and the energy eigenvalue is given by%
\begin{equation}
E=4\varepsilon +(2L+3)b.
\end{equation}%
The general solutions of the Schr\"{o}dinger equation with the sextic
oscillator potential can be obtained by substituting%
\begin{equation}
\psi (x)=x^{1+L}e^{-\frac{b}{2}x^{2}-\frac{q}{4}x^{4}}\Re _{j}(x^{2})
\end{equation}%
into the Schr\"{o}dinger equation(\ref{eq:5}). By rearranging the terms one
can obtain,%
\begin{equation}
\Re _{j}(x^{2})=\sum\limits_{m=0}^{2j}\frac{(2j)!(2L+1)!(L+m)!}{%
2m!(2j-m)!(2L+1+2m)!}P_{m}(\varepsilon )(4qa^{2}x^{2})^{m}
\end{equation}%
along with the recurrence relation satisfied by $P_{m}(\varepsilon )$%
\begin{equation}
2(2j-m)qa^{2}P_{m+1}(\varepsilon )+2(\varepsilon -bm)P_{m}(\varepsilon
)-m(2L+2m+1)P_{m-1}(\varepsilon )=0.
\end{equation}%
The first four polynomials are given by%
\begin{align}
P_{1}(\varepsilon )& ={}\varepsilon  \notag \\
P_{2}(\varepsilon )& ={}\varepsilon ^{2}-b\varepsilon -\frac{1}{2}%
(3+2L)qa^{2}  \notag \\
P_{3}(\varepsilon )& =\varepsilon ^{3}-3b\varepsilon
^{2}-2(b^{2}-2(L+2))\varepsilon +2b(3+2L)qa^{2}  \label{Eq:20} \\
P_{4}(\varepsilon )& ={}\varepsilon ^{4}-6b\varepsilon
^{3}+(11b^{2}-5(2L+5)qa^{2}\varepsilon
^{2}+3b(-2b^{2}+(10L+21)qa^{2}\varepsilon  \notag \\
& +9\left[ -b^{2}(2L+3)+\frac{1}{4}(4L(5L+1)+21)q^{2}a^{4}\right].  \notag
\end{align}

They agree with the polynomials given in references \cite{Bender1} and \cite%
{Krajewska}. We have also compared our results with the numerical solutions
obtained in the reference\cite{Daniel}, for the potentials, 
\begin{eqnarray}
V_{1}(x) &=&x^{2}+\frac{x^{4}}{2(7.625)^{3/2}}+\frac{x^{6}}{7442}  \notag \\
V_{2}(x) &=&x^{2}+\frac{x^{4}}{2(7.375)^{3/2}}+\frac{x^{6}}{6962} \\
V_{3}(x) &=&x^{2}+\frac{x^{4}}{2(7.125)^{3/2}}+\frac{x^{6}}{6498}  \notag
\end{eqnarray}%
where we have obtained exactly the same results of\cite{Daniel}, for the
energy eigenvalues%
\begin{equation}
E_{1}=2.897143,\quad E_{2}=5.891677,\quad and\quad E_{3}=8.991223.
\end{equation}%
corresponding to the potentials $V_{1}(x),V_{2}(x),and\ V_{3}(x)$%
respectively.

\section{PT symmetric anharmonic oscillator potential}

In order to transform the potential given in (\ref{eq:16}) to the PT
symmetric anharmonic oscillator potential we replace $q$ and $L$ by%
\begin{equation}
q\rightarrow \frac{q}{A^{2}},\quad L\rightarrow \frac{1}{8}\left(
qL^{2}+(2+4A)L-(20-32j)\right) .  \label{eq:19}
\end{equation}%
Further, substituting in the parameters in (\ref{eq:15}) and (\ref{eq:16})%
\begin{eqnarray}
A &=&\sqrt{17}\frac{qa^{2}}{\alpha ^{3}}+\frac{7}{\sqrt{17}}\frac{b}{\alpha
^{2}},\quad L=\frac{1}{2}\left( 3-\sqrt{17}+\left( 1-\frac{7}{\sqrt{17}}%
\right) \frac{\alpha b}{qa^{2}}\right)  \notag \\
\lambda &=&\frac{\varepsilon +2jb}{\alpha ^{2}}+\frac{4jqa^{2}}{\alpha ^{3}}%
,q\rightarrow \frac{4qa^{2}}{\alpha ^{3}}-\frac{1}{17}\left( 1+\frac{7}{%
\sqrt{17}}\right) \frac{2b^{2}+17\ell qa^{2}}{qa^{2}\alpha }  \label{eq:20}
\end{eqnarray}%
and taking the limit of the $(V(x)-E)$ given in (\ref{eq:15}) and(\ref{eq:16}%
) when $\alpha \rightarrow 0$, we obtain the PT symmetric anharmonic
oscillator potential:%
\begin{equation}
V=2i(b\ell -(1+2j)qa^{2})x+(b^{2}-2\ell
qa^{2})x^{2}+2iqba^{2}x^{3}-q^{2}a^{4}x^{4}.  \label{eq:21}
\end{equation}%
This leads to the energy eigenvalues,%
\begin{equation}
E=\varepsilon +b(1+2j)+\ell ^{2}.  \label{eq:22}
\end{equation}%
The ground state wave function of the potential can be obtained from (\ref%
{eq:17}) by using the same limiting procedure as introduced in (\ref{eq:19})
and (\ref{eq:20}):%
\begin{equation}
\psi =e^{(-ix-\frac{1}{2}bx^{2}-\frac{2i}{3}qA^{2}x^{3})}
\end{equation}%
The wave function for any $j$ can be obtained by letting%
\begin{equation}
\psi (x)=e^{(-ix-\frac{1}{2}bx^{2}-\frac{2i}{3}qA^{2}x^{3})}\sum%
\limits_{m=0}^{2j}a_{m}x^{2m}.
\end{equation}%
Here we obtain a four-term recurrence relation for the energy,%
\begin{equation}
2i(2j-m)qa^{2}P_{m+1}(\varepsilon )+(\varepsilon -2b(m-j))P_{m}(\varepsilon
)-2im\ell P_{m-1}(\varepsilon )+m(m-1)P_{m-2}(\varepsilon )=0
\end{equation}%
The first four $P_{m}$ is given by 
\begin{eqnarray}
P_{1}(\varepsilon ) &=&\varepsilon  \notag \\
P_{2}(\varepsilon ) &=&\varepsilon ^{2}-b^{2}-4qa^{2}\ell  \notag \\
P_{3}(\varepsilon ) &=&\varepsilon ^{3}-4(b^{2}+4qa^{2}\ell )\varepsilon
-16q^{2}a^{4} \\
P_{4}(\varepsilon ) &=&\varepsilon ^{4}-10(b^{2}+4qa^{2}l)\varepsilon
^{2}-96q^{2}a^{4}\varepsilon +9(b^{4}+8qb^{2}a^{2}\ell +16q^{2}a^{4}\ell
^{2})  \notag
\end{eqnarray}%
The results are in agreement with those given in\cite{Bender3}.

\section{Conclusions}

We have developed a general procedure to obtain the eigenstates of some QES
potentials in terms of the orthogonal polynomials as well as the
eigenvalues. We have proven that these potentials can be obtained from each
other by adequate transformations. Some examples have been presented to test
the validity of the procedure given here.

The present work can be generalized in various directions. The method we
have introduced can be applied to determine the QES forms of the other
exactly solvable potentials. The present construction seems to exhibit
closer similarity to the current perturbation constructions. In particular,
the perturbed P\"{o}schl-Teller potential can be worked out by following the
procedure discussed here and some of its eigenstates can be determined
exactly.


\begin{thebibliography}{9}
\bibitem{Turbiner1} Turbiner, A. V. and Ushveridze, A. G., Phys. Lett. A 
\textbf{126,} 181 (1987)

\bibitem{Turbiner2} Turbiner, A. V., Commun. Math. Phy. \textbf{118,} 467
(1988)

\bibitem{Gonzalez} Gonzalez-Lopez, A., Kamran, N., and Olver, P. J., Commun.
Math. Phys. \textbf{153}, 117 (1993)

\bibitem{Debergh} Debergh, N., J. Phys.A: \ Gen.Math. \textbf{33}, 7109
(2000)

\bibitem{Kuliy} Kuliy, T. V. and Thachuk, V. M., J. Phys.A: \ Gen.Math. 
\textbf{32}, 2157 (1999)

\bibitem{Brihaye} Brihaye, Y., Debergh, N. and Ndimubandi,  J., Mod. Phys.
Lett. A \textbf{16}, 1243 (2001)

\bibitem{Bender1} Bender, C. M. and Dunne, G. V., J. Math. Phys. \textbf{37,}
6 (1996)

\bibitem{Bender2} Bender C M, Dunne G V and Moshe M 1997 Phys. Rev. A 
\textbf{55,} 2625 (1997)

\bibitem{Finkel} Finkel F, Gonzalez-Lopez A and Rodiriguez M A 1996 \ J.
Math. Phys. \textbf{37, }3954 (1996)

\bibitem{Krajewska} Krajewska, A., Ushveridze, A. and Walczak, Z., Mod.
Phys. Lett. A\textbf{12,} 1131 (1997)

\bibitem{Znojil1} Znojil, M., Phys. Lett. A \textbf{266,} 254 (2000)

\bibitem{Znojil2} Znojil, M., 4$^{th}$Int.Conf. ''Symmetry in Nonlinear
Mathematical Physics'' July 9-15 2001 Kyiv Ukrania

\bibitem{Turbiner3} Turbiner, A.,  ''Handbook of Lie Group Analysis of
Differential equations V3: New Trends in Theoretical Developments and
Computational Methods'' (CRC\ Press ed. N. H. Ibragimov, 1995)

\bibitem{Bagchi} Bagchi, B. and Roychoudhury, R., J. Phys.A: \ Gen.Math. 
\textbf{33} L1 (2000)

\bibitem{Znojil3} Znojil, M., J. Phys.A: \ Gen.Math. \textbf{34} 9585 (2001)

\bibitem{Daniel} Morales, A. D. and Parra-Mejias, Z., Can. J. Phys. \textbf{%
77} 863 (1999)

\bibitem{Bender3} Bender, C. M. and Boettcher, S., J. Phys.A: \ Gen.Math. 
\textbf{31} L273 (1998)
\end{thebibliography}
\end{document}